\documentclass[aps,prapplied,reprint,superscriptaddress,longbibliography]{revtex4-2}

\usepackage{amsmath,amssymb}
\usepackage{multirow}
\usepackage{dsfont}
\usepackage{graphicx}
\usepackage{float}
\usepackage{dcolumn}
\usepackage{booktabs}
\usepackage{siunitx}
\usepackage{ulem}
\DeclareSIUnit{\torr}{Torr}

\newcommand{\dv}[2]{\frac{\mathrm{d}#1}{\mathrm{d}#2}}
\newcommand{\pdv}[2]{\frac{\partial #1}{\partial #2}}
\newcommand{\expval}[1]{\left\langle#1\right\rangle}
\newcommand{\vb}[1]{\mathbf{#1}}
\newcommand{\comm}[2]{\left[#1,#2\right]}
\newcommand{\abs}[1]{\left\lvert#1\right\rvert}

\usepackage{color}
\definecolor{mygreen}{rgb}{0,0.5,0} 
\definecolor{mygrey}{rgb}{0.5,0.5,0.5} 
\definecolor{myred}{rgb}{0.75,0,0} 
\definecolor{myblue}{rgb}{0,0,0.75} 
\definecolor{mymagenta}{cmyk}{0,1,0,0.12} 
\definecolor{mycyan}{cmyk}{1,0,0,0.12} 
\definecolor{myorange}{rgb}{0.85,0.375,0}  
\definecolor{myviolet}{rgb}{0.5,0.3,1} 
\definecolor{mybrown}{rgb}{0.542969,0.269531, 0.0742188}

\newcommand{\ICREAaddress}{ICREA---Instituci\'o Catalana de Recerca i Estudis Avan\c{c}ats, 08010 Barcelona, Spain}
\newcommand{\ICFOaddress}{ICFO---Institut de Ci\`encies Fot\`oniques, The Barcelona Institute of Science and Technology, 08860, Castelldefels (Barcelona), Spain}

\begin{document}

\title{Systematic errors from scalar slowing-down models in transient SERF magnetometry}

\author{Marek Kopciuch}
\email[Corresponding author: ]{Marek.Kopciuch@icfo.eu}
\affiliation{\ICFOaddress}

\author{Harini Raghavan}
\affiliation{\ICFOaddress}

\author{Morgan W. Mitchell}
\affiliation{\ICFOaddress}
\affiliation{\ICREAaddress}

\newcommand{\oldnew}[2]{{\color{red}old: \sout{#1}}{\color{blue}new: #2}}
\newcommand{\new}[1]{{\color{blue}{new: #1}}}
\newcommand{\old}[1]{{\color{red}{old: \sout{#1}}}}

\date{\today}

\begin{abstract}
Reduced Bloch models of spin-exchange-relaxation-free (SERF) magnetometers commonly describe the coupled electron--nuclear spin dynamics using a single nuclear slowing-down factor. We show that this scalar description becomes inaccurate when the polarization magnitude changes appreciably. We derive an effective Bloch equation with separate longitudinal and transverse slowing-down factors, $q_\parallel(P)$ and $q_\perp(P)$, governing changes in polarization magnitude and direction. For $^{87}\mathrm{Rb}$ under representative SERF conditions and undergoing free-induction decay (FID), comparison with master-equation simulations shows that the dynamic scalar model produces a magnetic-field estimation bias of up to $4.2\%$ at high initial polarization, while fixing the slowing-down factor at a reference value increases the error to $10.1\%$. Lowering the polarization suppresses this bias but also reduces the magnetic-field information contained in the FID signal. We further show that the anisotropic response modifies noncollinear optical pumping and introduces an additional nonlinear contribution to the harmonic response under periodic driving. The resulting model retains the simplicity of a three-component Bloch description while improving the quantitative accuracy of transient and driven SERF magnetometry.
\end{abstract}

\maketitle

\section{Introduction}

Sensors based on collective spin dynamics in atomic vapors underpin some of the most sensitive measurements of magnetic fields, frequency, and rotation. Optically pumped magnetometers (OPMs) infer magnetic fields from the precession and relaxation of spin polarization in alkali-metal vapors and now enable applications ranging from biomagnetic sensing, including wearable brain imaging \cite{Boto2018,Tierney2019}, to precision inertial sensing \cite{Wei2023}, atomic comagnetometry \cite{Limes2018}, zero- and ultralow-field nuclear magnetic resonance \cite{Bodenstedt2021} and searches for exotic spin-dependent interactions and ultralight dark matter \cite{Vasilakis2009,Afach2021,Zhang2026}. In all of these systems, the quantity of interest is encoded in the time evolution of a measured spin signal. The accuracy of that inference therefore depends directly on the dynamical model: an incomplete or oversimplified model can introduce a systematic bias that averaging cannot remove. Quantitatively accurate models of atomic spin dynamics are thus essential for interpreting the sensitivity of these devices.

The spin dynamics of an alkali-metal atom are governed by a master equation for its coupled electron–nuclear density matrix, including hyperfine coupling, magnetic precession, spin-exchange and spin-destruction collisions and optical pumping. Although exact within the adopted collision model, this description is computationally expensive and offers limited intuition for how an applied field appears in the observable signal. For practical modeling, it is therefore commonly reduced to a Bloch equation for the electron-spin polarization $\vb{P}$. In the spin-exchange-relaxation-free (SERF) regime, rapid spin-exchange collisions drive the atomic ensemble toward a spin-temperature state. Within this state, the electronic and nuclear spin polarizations are strongly coupled, so that the much more weakly field-responsive nuclear spin participates in the collective angular-momentum dynamics. As a result, the electron polarization responds more slowly to magnetic torques, conventionally described by a nuclear slowing-down factor $q(P)$.  This reduction underlies the original demonstrations of SERF magnetometry \cite{Allred2002,Kominis2003} and its extensions \cite{Ledbetter2008,Savukov2005}, and remains widely used for both steady-state and transient signals \cite{Tang2021,Padniuk2022,Mouloudakis2024}. Its accuracy, however, is rarely tested directly against the master equation. This is particularly concerning in the strongly polarized, transient regime relevant to pulsed and modulated SERF operation where both the magnitude and direction of the polarization change appreciably and therefore probe different differential responses of the spin-temperature state.

We derive the reduced Bloch equation directly from the master equation and identify explicitly where the scalar approximation enters. Retaining the full differential response of the spin-temperature manifold yields distinct longitudinal and transverse slowing-down factors, $q_{\parallel}(P)$ and $q_{\perp}(P)$, rather than a single $q(P)$. The transverse factor retains the usual interpretation of the nuclear slowing-down factor, governing changes in the polarization direction. The longitudinal factor instead governs changes in its magnitude, for which the nuclear polarization must readjust together with the electron spin. We then quantify the cost of the conventional scalar reduction for $^{87}\mathrm{Rb}$ under representative SERF conditions, given in Appendix~\ref{app:common_simulation_parameters} and based on the vapor cell of Ref.~\cite{Raghavan2024}. We find that the dynamic scalar model produces a magnetic-field estimation bias of up to $4.2\%$ at high polarization, increasing to $10.1\%$ when the slowing-down factor is frozen at a fixed reference value. We further show that operating at lower polarization can suppress this bias, but only at the cost of sacrificing most of the magnetic-field information contained in the transient. We trace the error to the underlying anisotropy of the spin-temperature response: applying a single slowing-down factor to both polarization-magnitude relaxation and directional precession produces an incorrect longitudinal trajectory $P(t)$,  which in turn biases the accumulated precession angle and the inferred magnetic field. Our results therefore provide a quantitative basis for retaining separate longitudinal and transverse slowing-down factors when modeling transient and driven SERF magnetometry. While this work was being prepared, Koutrouli, Vasilakis, and Mouloudakis independently reported a complementary microscopic derivation of distinct longitudinal and transverse slowing-down factors \cite{Koutrouli2026}. Their work focuses on the microscopic origin of these factors.

In Sec.~\ref{sec:model} we derive an anisotropic effective Bloch equation from the ground-state master equation and identify the hierarchy connecting it to dynamic and fixed scalar-$q$ models. In Sec.~\ref{sec:relaxation} we show that the scalar approximation modifies the longitudinal free-polarization decay and, through the resulting trajectory $P(t)$, the accumulated precession angle. Section~\ref{sec:field-estimation-bias} quantifies the corresponding magnetic-field bias and its trade-off with available Fisher information, while Sec.~\ref{sec:phase-envelope} derives closed-form amplitude--precession-angle relations that isolate the anisotropic contribution analytically. Section~\ref{sec:pumping} considers noncollinear optical pumping and shows that the anisotropic longitudinal pumping correction provides an additional nonlinear channel for high-harmonic generation under periodic driving.

\section{Effective Bloch model and hierarchy of approximations}
\label{sec:model}

We consider the ground-state dynamics of an alkali-metal atom with electron spin $S=1/2$ in the high-buffer-gas-pressure regime relevant to the SERF cell modeled below \cite{Raghavan2024}. In this regime, excited-state collisions suppress atomic alignment, so that optical pumping can be described in terms of spin orientation alone. We start from the master equation \cite{Happer1977,Appelt1998}:
\begin{equation}
\begin{split}
    \dv{\rho}{t}
    &=
    -i\comm{\widehat{\mathcal{H}}(t)}{\rho}
    +
    R_{\mathrm{SD}}\left[\varphi(\rho)-\rho\right]\\
    &\quad+
    R_{\mathrm{SE}}
    \left[
    \varphi(\rho)
    \left(
    \widehat{\mathds{1}}+4\expval{\widehat{\vb{S}}}\cdot\widehat{\vb{S}}
    \right)
    -\rho
    \right]\\
    &\quad +
    R_{\mathrm{OP}}
    \left[
    \varphi(\rho)
    \left(
    \widehat{\mathds{1}}+2\vb{s}\cdot\widehat{\vb{S}}
    \right)
    -\rho
    \right],
    \label{eq:master_serf}
\end{split}
\end{equation}
where $R_{\mathrm{SD}}$, $R_{\mathrm{SE}}$, and $R_{\mathrm{OP}}$ are the spin-destruction, spin-exchange, and optical-pumping rates, respectively. The dimensionless pumping vector $\vb{s}$ lies along the pump propagation axis, with its orientation set by the helicity and its magnitude by the degree of circular polarization; $\abs{\vb{s}}=1$ corresponds to ideal circular pumping. For $S=1/2$, the map $\varphi(\rho)=(\widehat{\mathds{1}}_S/2)\otimes\operatorname{Tr}_S\rho$ randomizes the electron spin while preserving the reduced nuclear state. The Hamiltonian, expressed in angular-frequency units, is
\begin{equation}
    \widehat{\mathcal{H}}(t)
    =
    \omega_{\mathrm{hf}}\,\widehat{\vb{I}}\cdot\widehat{\vb{S}}
    +
    \gamma_e\,\vb{B}(t)\cdot\widehat{\vb{S}}
    +
    \gamma_I\,\vb{B}(t)\cdot\widehat{\vb{I}},
    \label{eq:hamiltonian_serf}
\end{equation}
where $\widehat{\vb{I}}$ and $\widehat{\vb{S}}$ are the dimensionless nuclear and electronic spin operators, $\omega_{\mathrm{hf}}$ is the hyperfine coupling constant, $\vb{B}(t)$ is the applied magnetic field, and $\gamma_e$ and $\gamma_I$ are the signed electronic and nuclear gyromagnetic coefficients. Following the derivation in Appendix~\ref{app:bloch_derivation}, we add the electronic and nuclear first-moment equations for the total angular momentum $\widehat{\vb{F}}=\widehat{\vb{I}}+\widehat{\vb{S}}$. The internal hyperfine torques cancel, giving
\begin{equation}
\begin{split}
\dv{}{t}\expval{\widehat{\vb{F}}}
&=
\gamma_e\vb{B}(t)\times\expval{\widehat{\vb{S}}}
+\gamma_I\vb{B}(t)\times\expval{\widehat{\vb{I}}}\\
&\quad-R_{\mathrm{SD}}\expval{\widehat{\vb{S}}}
+R_{\mathrm{OP}}\left(\frac{\vb{s}}{2}-\expval{\widehat{\vb{S}}}\right).
\end{split}
\label{eq:F_exact}
\end{equation}
Equation~\eqref{eq:F_exact} is exact within the adopted master-equation model, but is not yet closed in terms of the electron polarization $\vb{P}=2\expval{\widehat{\vb{S}}}$. 

Our first approximation assumes that rapid spin exchange keeps the atomic state close to the spin-temperature manifold while magnetic precession, relaxation, pumping, and external modulation evolve on slower time scales \cite{Happer1977,Appelt1998}:
\begin{equation}
    \rho_{\mathrm{ST}}
    =\frac{\exp\left(\beta\hat{\vb{n}}\cdot\widehat{\vb{F}}\right)}{Z},
    \qquad
    P=\tanh\left(\frac{\beta}{2}\right).
    \label{eq:spin_temperature}
\end{equation}
Here $P=\abs{\vb{P}}$, $\hat{\vb{n}}=\vb{P}/P$ is the polarization direction, $\beta$ controls its magnitude, and $Z$ normalizes the density matrix. On this manifold the electronic and nuclear first moments are collinear, and for $I=3/2$, as in $^{87}\mathrm{Rb}$, the magnitude of the mean total angular momentum is
\begin{equation}
F(P)
=
\abs{\expval{\widehat{\vb{F}}}}
=
\frac{3+P^2}{1+P^2}P,
\label{eq:F_magnitude}
\end{equation}
Thus $\expval{\widehat{\vb{F}}}=F(P)\hat{\vb{n}}$ and $\vb{P}=P\hat{\vb{n}}$. The nonlinearity of $F(P)$ produces the slowing-down anisotropy: changes in magnitude and direction probe different coefficients of this map, as seen directly from the chain rule,
\begin{align}
\dv{}{t}\expval{\widehat{\vb{F}}}
&=
\dv{F}{P}\dv{P}{t}\hat{\vb{n}}
+
F(P)\dv{\hat{\vb{n}}}{t},
\label{eq:F_chain_rule}\\
\dv{\vb{P}}{t}
&=
\dv{P}{t}\hat{\vb{n}}
+
P\dv{\hat{\vb{n}}}{t}.
\label{eq:P_chain_rule}
\end{align}
The first terms describe changes in the magnitude of the polarization, whereas the second describe changes in its direction. Splitting Eq.~\eqref{eq:F_exact} accordingly gives the transverse and longitudinal nuclear slowing-down factors 
\begin{equation}
        q_\perp(P)=2\frac{F(P)}{P},
        \qquad
        q_\parallel(P)=2\dv{F}{P}.
        \label{eq:slowing_factors}
\end{equation}

For $^{87}\mathrm{Rb}$ they coincide at $P=0$, where $q_\perp=q_\parallel=6$, but reach $q_\perp=4$ and $q_\parallel=2$ at full polarization. The same construction applies to other nuclear spins (see Fig.~\ref{fig:slowing_factors} and Tab.~\ref{tab:slowing_factors}). Substituting the spin-temperature moments into Eq.~\eqref{eq:F_exact} gives the closed anisotropic Bloch equation for $^{87}\mathrm{Rb}$,

\begin{align}
\dv{\vb{P}}{t}={}&\frac{1}{q_\perp(P)}\left\{
\left[\gamma_e+\gamma_I\frac{5+P^2}{1+P^2}\right]
\vb{B}(t)\times\vb{P}+R_{\mathrm{OP}}\vb{s}_\perp\right\}
\nonumber\\
&+\frac{1}{q_\parallel(P)}\left[
-R_{\mathrm{SD}}\vb{P}+R_{\mathrm{OP}}\left(\vb{s}_\parallel-\vb{P}\right)\right],
\label{eq:anisotropic_bloch}
\end{align}
where $\vb{s}_\parallel$ and $\vb{s}_\perp$ are the components of the pumping vector parallel and perpendicular to the instantaneous polarization,
\begin{equation}
    \vb{s}_\parallel \equiv (\vb{s}\cdot\hat{\vb{n}})\hat{\vb{n}},
    \qquad
    \vb{s}_\perp \equiv \vb{s}-\vb{s}_\parallel.
    \label{eq:pump_decomposition}
\end{equation}
Magnetic precession is purely transverse, spin destruction is purely longitudinal, and optical pumping generally drives both components. At $P=0$ the two factors coincide, so Eq.~\eqref{eq:anisotropic_bloch} has a continuous isotropic limit independent of the choice of $\hat{\vb{n}}$.

The nonlinear equation generally requires numerical integration, motivating a hierarchy of further approximations. First, the nuclear magnetic moment is much smaller than the electronic one, $\abs{\gamma_I}\ll\abs{\gamma_e}$, allowing
\begin{equation}
    \gamma_e+\gamma_I\frac{5+P^2}{1+P^2} \approx \gamma_e.
    \label{eq:neglect_nuclear_zeeman}
\end{equation}
This removes a small correction to precession while preserving the slowing-down anisotropy.

The scalar approximation conventionally used in optical-pumping and SERF
descriptions originates from considering a polarized ensemble whose evolution
predominantly rotates $\vb{P}$ rather than changing its magnitude
\cite{Appelt1998}. Setting $\dv{}{t}P=0$ removes the longitudinal response,
leaving the single transverse factor $q_\perp(P)=2F(P)/P$. The scalar model
then applies this factor to the entire equation, defining
$q(P)\equiv q_\perp(P)$.
After neglecting the nuclear Zeeman term, this gives the dynamic scalar model
\begin{equation}
    \dv{\vb{P}}{t}=\frac{1}{q(P)}\Big[\gamma_e\vb{B}(t)\times\vb{P}-R_{\mathrm{SD}}\vb{P}+R_{\mathrm{OP}}\left(\vb{s}-\vb{P}\right)\Big].
\label{eq:dynamic_scalar_bloch}
\end{equation}

Finally, the fixed scalar model freezes the factor at a reference polarization,
\begin{equation}
    q\big(P(t)\big)\longrightarrow q_{\mathrm{ref}}\equiv q(P_{\mathrm{ref}}),
    \label{eq:fixed_scalar_q}
\end{equation}
chosen, for example, at a steady-state operating point or in the low-polarization limit. This assumption can enter implicitly when a transient is modeled by a purely exponential envelope. Comparing the anisotropic and dynamic scalar models isolates the effect of slowing-down anisotropy; the fixed scalar model adds the error associated with ignoring the polarization dependence of $q$.

\section{Pump-free transient dynamics}
\label{sec:relaxation}

We now consider free polarization decay after the optical pump is switched off, $R_{\mathrm{OP}}=0$, with $\vb{P}(0)=P_0\hat{\vb{z}}$ and a static transverse magnetic field $\vb{B}=B\hat{\vb{x}}$. Initially neglecting the nuclear Zeeman correction as in Eq.~\eqref{eq:neglect_nuclear_zeeman}, the trajectory remains in the $yz$ plane and can be parameterized as
\begin{equation}
\begin{split}
    \vb{P}(t)&=P(t)\hat{\vb{u}}[\phi(t)],\\
    \hat{\vb{u}}(\phi)&=\cos\phi\,\hat{\vb{z}}-\sin\phi\,\hat{\vb{y}}.
\end{split}
\label{eq:fid_trajectory}
\end{equation}
Here $P(t)=\abs{\vb{P}(t)}$ is the polarization magnitude and $\phi(t)$ is the continuously accumulated precession angle, with $\phi(0)=0$. Projecting the Bloch equation onto the radial and angular directions gives
\begin{align}
    \dv{P}{t}&=-\frac{R_{\mathrm{SD}}}{q_\parallel \big(P(t)\big)}P(t),
    \label{eq:fid_amplitude}\\
    \dv{\phi}{t}&=\frac{\gamma_e B}{q_\perp \big(P(t) \big)}.
    \label{eq:fid_phase}
\end{align}
The derivation is given in Appendix~\ref{app:fid_derivation}. These equations separate the roles of the two slowing-down factors: $q_\parallel$ controls the decay of the envelope, whereas $q_\perp$ controls the instantaneous precession frequency. The amplitude evolves independently of the precession angle, but the precession angle rate follows the changing amplitude through $q_\perp \big(P(t) \big)$.

\begin{figure}[t]
    \centering
    \includegraphics[width=\columnwidth]{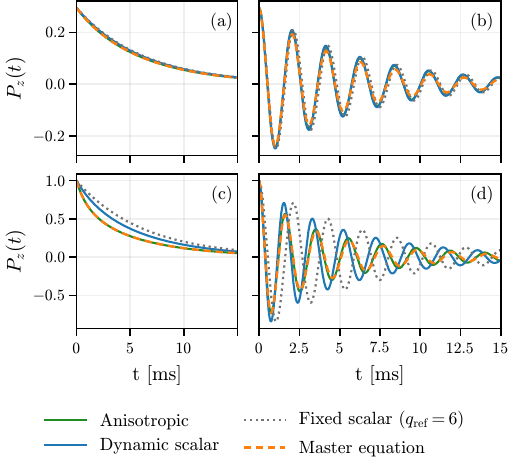}
    \caption{Free polarization decay for initial polarizations $P_0\simeq0.3$ (top row) and $1$ (bottom row), with $B=\qty{0}{\nano\tesla}$ (left column) and \qty{100}{\nano\tesla} (right column). At $t=0$ the pump is switched off and the indicated field is applied. The master-equation reference is compared with the anisotropic, dynamic scalar, and fixed scalar ($q_{\mathrm{ref}}=6$) models over a \qty{15}{\milli\second} interval. All Bloch curves neglect the nuclear Zeeman correction. Simulation details are listed in Appendix~\ref{app:numerics}.
    }
    \label{fig:relaxation_comparison}
\end{figure}

Because Eq.~\eqref{eq:fid_amplitude} does not depend on $\phi$, the amplitude dynamics are the same with or without transverse precession. Setting $B=0$ therefore provides a convenient way to display the relaxation law in isolation, but is not required for its solution. For constant $R_{\rm SD}$, Eq.~\eqref{eq:fid_amplitude} formally integrates to
\begin{equation}
    P(t)=P_0\exp\left[-\int_0^t
    \frac{R_{\mathrm{SD}}}{q_\parallel \big( P(t') \big)}\,\mathrm{d}t'\right].
    \label{eq:decay_rate_history}
\end{equation}
Writing the solution in this form does not make the decay exponential: a purely exponential envelope requires a constant logarithmic rate. Both the anisotropic and dynamic scalar models retain a polarization-dependent slowing-down factor, so their decay is generally nonexponential. A constant factor $q = q_\mathrm{ref}$ gives the exponential baseline $P(t)=P_0\exp(-R_{\mathrm{SD}}t/q_{\mathrm{ref}})$. The departure from exponential relaxation becomes particularly visible at high initial polarization, as illustrated by comparing panels (a) and (c) of Fig.~\ref{fig:relaxation_comparison}.

Figure~\ref{fig:relaxation_comparison} compares the master-equation reference with the three reduced models for two initial polarizations, at zero field and in a transverse field. This and the following simulations use the vapor-cell parameters of Ref. \cite{Raghavan2024}, listed explicitly in Table \ref{tab:common_simulation_parameters} of Appendix~\ref{app:common_simulation_parameters}. All models are initialized with the same electronic polarization extracted from the prepared master-equation state. The preparation protocol and numerical parameters are listed in Appendix~\ref{app:numerics}. The fixed scalar baseline uses $q_{\mathrm{ref}}=q_{\perp}(0)=6$, corresponding to the unpolarized steady state reached with the pump switched off. For the strongly polarized states, the anisotropic model follows the reference substantially more closely than either scalar approximation.

Although Eq.~\eqref{eq:fid_phase} contains no explicit dependence on $q_\parallel$, slowing-down anisotropy still modifies the accumulated precession angle indirectly. At low polarization, where $q_\parallel \approx q_\perp$, the models predict nearly identical amplitude decay [Fig.~\ref{fig:relaxation_comparison}(a)], and their precession angles therefore remain closely aligned [Fig.~\ref{fig:relaxation_comparison}(b)]. At high polarization, the increasing separation between $q_\parallel$ and $q_\perp$ produces distinct decay trajectories [Fig.~\ref{fig:relaxation_comparison}(c)], which in turn lead to a clear divergence of the predicted precession [Fig.~\ref{fig:relaxation_comparison}(d)].

\section{Magnetic-field estimation bias}
\label{sec:field-estimation-bias}

Equation~\eqref{eq:fid_phase} shows that the static magnetic field enters this pump-free Bloch evolution through the accumulated precession angle, while the amplitude equation is independent of the field. The angle therefore carries the magnetic-field information in this description. For $\phi(0)=0$, integration gives
\begin{equation}
    \phi(t)=\gamma_e B\tau(t), \qquad
    \tau(t)=\int_0^t\frac{\mathrm{d}t'}{q_\perp \big( P(t') \big)}.
    \label{eq:fid_phase_effective_time}
\end{equation}
Eq.~\eqref{eq:fid_phase_effective_time} shows that the accumulated precession angle remains linear in $B$, with the conversion between field and angle determined by $\gamma_e\tau(t)$. Here $\tau(t)$ is the laboratory time accumulated at the reduced rate $1/q_\perp\big(P(t)\big)$, thereby incorporating the time-dependent slowing of the precession. Since this quantity has units of time, we refer to it as the effective time.

Equation~\eqref{eq:fid_phase_effective_time} applies to the Bloch equation with the nuclear Zeeman term neglected. To retain the same form of the relation when this term is included, the effective time for $^{87}\mathrm{Rb}$ becomes
\begin{equation}
    \tau_{\mathrm{eff}}(t)=\int_0^t \frac{1}{q_\perp \big(P(t')\big)}
    \left[ 1+\dfrac{\gamma_I}{\gamma_e}
    \dfrac{5+P(t')^2}{1+P(t')^2} \right]\mathrm{d}t'.
    \label{eq:fid_effective_time_nuclear}
\end{equation}

\begin{figure}[!htbp]
    \centering
    \includegraphics[width=\columnwidth]{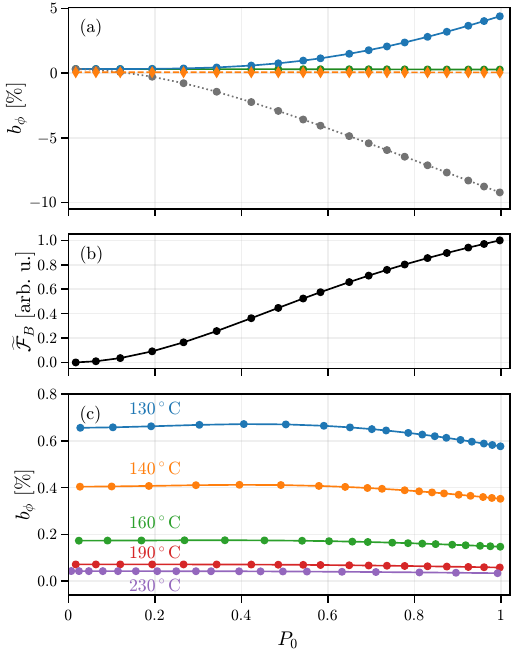}
    \caption{Precession angle bias and magnetic-field information versus initial polarization. (a) Relative accumulated-precession bias $b_\phi$ for the anisotropic, dynamic scalar, and fixed scalar ($q=6$) models, with colors as in Fig.~\ref{fig:relaxation_comparison}. The additional dashed curve includes the nuclear Zeeman correction in the anisotropic model. (b) Master-equation Fisher information normalized to its maximum. (c) Residual bias of the nuclear-corrected anisotropic model at the temperatures indicated in the boxes. A transverse field of \qty{5}{\nano\tesla} is applied after preparation. Simulation details are listed in Appendix~\ref{app:numerics}.}
    \label{fig:fid_phase_bias_and_fisher_vs_initial_polarization}
\end{figure}

In this description, the precession angle bias predicted by a dynamical model translates directly into a systematic error of the magnetic-field estimate. We define the fractional bias
\begin{equation}
    b_\phi=
    \frac{\phi_{\mathrm{model}}-\phi_{\mathrm{ME}}}
    {\phi_{\mathrm{ME}}},
    \label{eq:fid_phase_bias}
\end{equation}
where $\phi_{\mathrm{model}}$ and $\phi_{\mathrm{ME}}$ are accumulated over the same FPD interval at the same reference field, with $\phi_{\mathrm{model}}(0)=\phi_{\mathrm{ME}}(0)=0$. Inferring $B$ from the measured accumulated precession angle using the biased, model-predicted $\tau$ gives the fractional field bias
\begin{equation}
    \frac{\overline{B}-B}{B}
    =-\frac{b_\phi}{1+b_\phi}\simeq-b_\phi,
    \label{eq:fid_field_bias}
\end{equation}
where the last approximation assumes $\abs{b_\phi}\ll1$. A model that overpredicts the accumulated precession therefore underestimates the field, and conversely.

Figure~\ref{fig:fid_phase_bias_and_fisher_vs_initial_polarization}(a) shows a common positive precession angle bias of about $0.3\%$ for the three electronic-only Bloch models at low initial polarization. At high polarization, the dynamic scalar model develops an increasingly positive bias, reaching about $4.4\%$, while the fixed scalar model underpredicts the accumulated precession by about $9.2\%$. The anisotropic model remains close to a $0.3\%$ bias throughout the polarization range. By Eq.~\eqref{eq:fid_field_bias}, these precession-angle biases correspond to magnetic-field biases of approximately $4.2\%$ for the dynamic scalar model and $10.1\%$ for the fixed scalar model, whereas the anisotropic model remains near $0.3\%$.

Reducing the initial polarization might therefore appear to offer a simple way to suppress the error of the dynamic scalar model: around $P_0\simeq0.3$, its bias becomes comparable to that of the anisotropic model. However, the smaller signal amplitude also reduces the precision of precession angle estimation. For a fixed additive polarization-noise level, the local angle uncertainty scales as $\sigma_\phi(t)\propto\sigma_P/P(t)$, so a larger amplitude at a given time allows a more precise precession measurement.

To quantify this cost, we use the Fisher information, which measures how much information about $B$ is contained in the measured FPD trajectory. Figure~\ref{fig:fid_phase_bias_and_fisher_vs_initial_polarization}(b) reports it relative to the largest value obtained across the $P_0$ scan; its full definition and the assumed noise model are given in Appendix~\ref{app:numerics}. Reducing $P_0$ to approximately $0.3$ leaves only about $20\%$ of the maximum Fisher information, reached for the nearly fully polarized gas. Thus, the reduction of scalar-model bias comes at the cost of discarding about $80\%$ of the available information in this measurement model.

The remaining bias of the electronic-only anisotropic model is approximately $0.3\%$. Retaining the nuclear Zeeman correction at the reference temperature reduces the bias considerably.  For example, at $T=\qty{190}{\degreeCelsius}$, it drops to approximately $0.06$--$0.07\%$, depending on $P_0$.  We attribute the smaller residual to finite-rate departures from the spin-temperature state assumed in the Bloch closure. Figure~\ref{fig:fid_phase_bias_and_fisher_vs_initial_polarization}(c) supports this interpretation: with the cell parameters, preparation protocol, and applied field held fixed, the residual bias of the nuclear-corrected anisotropic model decreases from about $0.6\%$ at \qty{130}{\degreeCelsius} to about $0.04\%$ at \qty{230}{\degreeCelsius}. Each temperature is compared with its own master-equation reference and amplitude cutoff. Under these conditions, changing temperature primarily modifies the collision rates through the equilibrium rubidium density; both spin-exchange and spin-destruction rates change. Spin exchange becomes faster relative to precession and spin destruction as the temperature rises, improving the separation of time scales underlying the SERF spin-temperature approximation.

\section{Amplitude--precession relation}
\label{sec:phase-envelope}

We now rederive the precession angle error identified in Secs.~\ref{sec:relaxation} and~\ref{sec:field-estimation-bias} analytically by replacing time with the polarization amplitude $P$ as the trajectory variable, obtaining closed-form relations between the polarization amplitude and accumulated precession angle. We retain the pump-free, constant-field geometry and initially neglect the nuclear Zeeman term. Because $P$ decreases monotonically, it can parameterize the trajectory, and taking the ratio of Eqs.~\eqref{eq:fid_amplitude} and~\eqref{eq:fid_phase} gives
\begin{equation}
    \dv{\phi}{P}
    =-\frac{\gamma_e B}{R_{\mathrm{SD}}}
    \frac{q_\parallel(P)}{q_\perp(P)}\frac{1}{P}.
    \label{eq:phase_envelope_derivative}
\end{equation}
The precession angle accumulated as the polarization decreases from $P_0$ to $P$ therefore depends on the slowing-down factors only through the ratio $q_{\parallel}/q_{\perp}$, with the time parametrization of the decay eliminated.

For $^{87}\mathrm{Rb}$, the ratio is $(3+P^4)/[(1+P^2)(3+P^2)]$, which by integrating Eq.~\eqref{eq:phase_envelope_derivative} with $\phi(P_0)=0$ yields the anisotropic-model precession angle
\begin{equation}
    \phi_{\mathrm{an}}(P)
    =\frac{\gamma_e B}{R_{\mathrm{SD}}}
    \ln\!\left[
    \frac{P_0(P_0^2+3)(1+P^2)}
    {P(P^2+3)(1+P_0^2)}
    \right].
    \label{eq:phase_envelope_anisotropic}
\end{equation}
In either scalar model, the same slowing-down factor multiplies the amplitude and precession angle dynamics and cancels from their ratio. Hence
\begin{equation}
    \phi_{\mathrm{sc}}(P)
    =\frac{\gamma_e B}{R_{\mathrm{SD}}}
    \ln\!\left(\frac{P_0}{P}\right).
    \label{eq:phase_envelope_scalar}
\end{equation}
This result holds for both a polarization-dependent scalar factor and a fixed factor. Starting from the same $P_0$, the dynamic and fixed scalar models predict the same accumulated precession when their amplitudes reach a common prescribed value $P$; they differ in the time required to reach that value.

This equality is consistent with the different scalar-model precession biases in Fig.~\ref{fig:fid_phase_bias_and_fisher_vs_initial_polarization}(a). There, for each initial state, the master-equation trajectory sets a common analysis endpoint for all models. The scalar trajectories generally have different amplitudes at that endpoint, so Eq.~\eqref{eq:phase_envelope_scalar} is evaluated at different values of $P$. Comparing accumulated precession angles at a common polarization amplitude instead allows each model to reach that amplitude at its own time. The two comparisons therefore impose different endpoints on the trajectories.

At a common amplitude the anisotropic correction to the scalar precession is
\begin{equation}
    \Delta\phi_{\mathrm{an}}(P)
    =\frac{\gamma_e B}{R_{\mathrm{SD}}}
    \ln\!\left[
    \frac{(P_0^2+3)(1+P^2)}
    {(1+P_0^2)(P^2+3)}
    \right].
    \label{eq:phase_envelope_anisotropy_correction}
\end{equation}
For $\gamma_e B>0$ and $0<P<P_0\leq1$, this correction is negative: the anisotropic model accumulates less precession than either scalar model before reaching the same amplitude.

For a known $R_{\mathrm{SD}}$, interpreting an anisotropic FPD precession angle with Eq.~\eqref{eq:phase_envelope_scalar} therefore yields $\overline{B}<B$. This systematic underestimation holds for both fixed and dynamic scalar models throughout $0<P<P_0\leq1$ when the precession angle is measured up to a prescribed polarization amplitude within the adopted pump-free model.

Retaining the nuclear Zeeman torque in Eq.~\eqref{eq:fid_frequency_appendix} leaves the anisotropic amplitude unchanged and adds the angle correction
\begin{equation}
\begin{split}
    \Delta\phi_{\mathrm{nuc}}(P)
    ={}&\frac{\gamma_I B}{R_{\mathrm{SD}}}
    \int_P^{P_0}\frac{5+p^2}{1+p^2}
    \frac{q_\parallel(p)}{q_\perp(p)}\frac{\mathrm{d}p}{p}.
\end{split}
    \label{eq:phase_envelope_nuclear_correction}
\end{equation}
Thus $\phi_{\mathrm{an+nuc}}=\phi_{\mathrm{an}}+\Delta\phi_{\mathrm{nuc}}$. For $0<P<P_0$, the integral is positive. With $B>0$ and $R_{\mathrm{SD}}>0$, the correction is therefore negative because $\gamma_I<0$ in our $^{87}\mathrm{Rb}$ convention.

\begin{figure}[t]
    \centering
    \includegraphics[width=\columnwidth]{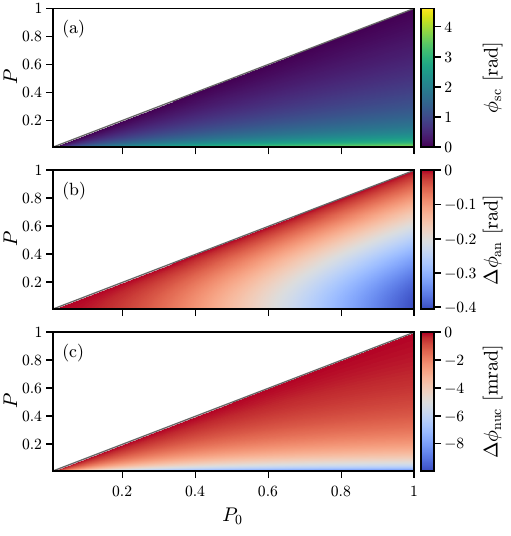}
    \caption{Analytical amplitude-precession maps for $^{87}\mathrm{Rb}$ at $\gamma_e B_\perp/R_{\mathrm{SD}}=1$. (a) Precession angle $\phi_{\mathrm{sc}}$ shared by the dynamic and fixed scalar models. (b) Anisotropy correction $\Delta\phi_{\mathrm{an}}=\phi_{\mathrm{an}}-\phi_{\mathrm{sc}}$, where $\phi_{\mathrm{an}}$ is the anisotropic-model precession. (c) Nuclear Zeeman correction $\Delta\phi_{\mathrm{nuc}}=\phi_{\mathrm{an+nuc}}-\phi_{\mathrm{an}}$. To keep the color scales readable, both $P$ and $P_0$ are sampled between $0.01$ and $1$, retaining only $P\leq P_0$.}
    \label{fig:phase_envelope_maps}
\end{figure}

Figure~\ref{fig:phase_envelope_maps} separates the scalar precession angle from the anisotropic and nuclear corrections at a common amplitude. The anisotropic correction grows in magnitude as a highly polarized initial state decays. At the highest total precession, $P_0=1$ and $P=0.01$, the scalar precession angle is approximately \qty{4.61}{\radian}, the anisotropic correction is \qty{-0.405}{\radian}, and the nuclear correction is \qty{-9.96}{\milli\radian}.

The maps use $\gamma_e B/R_{\mathrm{SD}}=1$. In this normalization, the nuclear prefactor is $\gamma_I B/R_{\mathrm{SD}}=\gamma_I/\gamma_e\simeq-4.970\times10^{-4}$. For the cell parameters in Appendix~\ref{app:numerics} at $T=\qty{190}{\degreeCelsius}$, the calculated rate is $R_{\mathrm{SD}}\simeq\qty{961}{\per\second}$, so this scale corresponds to $B=R_{\mathrm{SD}}/\gamma_e\simeq\qty{5.46}{\nano\tesla}$. Equations~\eqref{eq:phase_envelope_anisotropic}--\eqref{eq:phase_envelope_nuclear_correction} show that, for fixed $P$, $P_0$, and isotope, both the accumulated precession angle and its corrections scale linearly with $\gamma_e B/R_{\mathrm{SD}}$. The values shown in Fig.~\ref{fig:phase_envelope_maps} can therefore be multiplied by this ratio for another field or spin-destruction rate within the same pump-free model.

\section{Noncollinear optical pumping and driven configurations}
\label{sec:pumping}

We now consider optical pumping when the pumping vector $\vb{s}$ is not collinear with the instantaneous atomic polarization $P$. The optical-pumping contribution to the anisotropic Bloch equation, Eq.~\eqref{eq:anisotropic_bloch}, reads
\begin{equation}
    \left.\dv{\vb{P}}{t}\right|_{\mathrm{OP}}
    ={}-\frac{R_{\mathrm{OP}}}{q_\parallel(P)}\vb{P}
    +
    \underbrace{
    R_{\mathrm{OP}}\left[
    \frac{\vb{s}_\perp}{q_\perp(P)}
    +\frac{\vb{s}_\parallel}{q_\parallel(P)}
    \right]
    }_{\displaystyle \vb{G}}. \label{eq:anisotropic_optical_pumping}
\end{equation}
The anisotropy in Eq.~\eqref{eq:anisotropic_optical_pumping} enters both the first term, which describes repumping loss of the existing electron polarization, and the second term, which is the pumping source determined by the photon-spin polarization $\vb{s}$. It is therefore useful to examine the source term $\vb{G}$ first, but a physically complete comparison with the scalar model must include both contributions.

It is obtained by projecting $\vb{s}$ parallel and perpendicular to the instantaneous polarization, as in Eq.~\eqref{eq:pump_decomposition}, and recombining these components with different slowing-down weights. Unlike the scalar source $\vb{G}_{\mathrm{sc}}=R_{\mathrm{OP}}\vb{s}/q_\perp(P)$, this can change the source direction when both projections are nonzero and $q_\parallel\neq q_\perp$:
\begin{equation}
\begin{split}
    \vb{G}
    ={}&R_{\mathrm{OP}}\left[
    \frac{\vb{s}_\perp}{q_\perp(P)}
    +\frac{\vb{s}_\parallel}{q_\parallel(P)}
    \right]\\
    ={}&\vb{G}_{\mathrm{sc}}
    +R_{\mathrm{OP}}\left[\frac{1}{q_\parallel(P)}-\frac{1}{q_\perp(P)}\right]
    (\vb{s}\cdot\hat{\vb{n}})\hat{\vb{n}}.
\end{split}   \label{eq:anisotropic_optical_pumping_source}
\end{equation}
Because the two projections are rescaled differently, the effective source need not remain parallel to the photon-spin polarization.

\begin{figure}[t]
    \centering
    \includegraphics[width=\columnwidth]{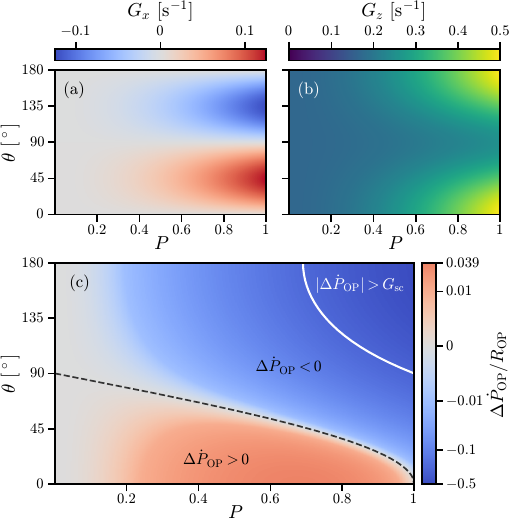}
    \caption{Optical-pumping source and amplitude-dynamics correction for $^{87}\mathrm{Rb}$ at $R_{\mathrm{OP}}=\qty{1}{\per\second}$, versus the polarization amplitude $P$ and angle $\theta$ relative to the pump. (a) $G_{x}$, generated by the slowing-down anisotropy. (b) $G_{z}$. Their color scales are symmetric about zero for $x$ and start at zero for $z$. (c) Full pumping correction relative to the dynamic scalar model, including repumping. The colors retain a symmetric normalization, linear within $\pm0.01$ and logarithmic outside, while the colorbar shows only the data range. The dashed curve marks zero correction; the white curve marks $\abs{\Delta\dot{P}_{\mathrm{OP}}}=G_{\mathrm{sc}}$.}
    \label{fig:optical_pumping_drive_maps}
\end{figure}

Without loss of generality, we choose the pump axis along $\hat{\vb{z}}$ and the polarization in the $xz$ plane, tilted toward $+\hat{\vb{x}}$. For $\vb{s}=\hat{\vb{z}}$, we write $\hat{\vb{n}}(\theta)=\cos\theta\,\hat{\vb{z}}+\sin\theta\,\hat{\vb{x}}$, where $0\leq\theta\leq\pi$ is the angle between the polarization and the pump. The source components in the plane spanned by the pump axis and the polarization are then
\begin{equation}
\begin{aligned}
    G_{x}
    &=\frac{R_{\mathrm{OP}}}{2}\sin(2\theta)
    \left[\frac{1}{q_\parallel(P)}-\frac{1}{q_\perp(P)}\right],\\
    G_{z}
    &=R_{\mathrm{OP}}\left[
    \frac{\sin^2\theta}{q_\perp(P)}+\frac{\cos^2\theta}{q_\parallel(P)}
    \right].
\end{aligned}
    \label{eq:optical_pumping_source_components}
\end{equation}
The transverse source component $G_x$ vanishes for collinear and orthogonal pump--polarization geometries, is maximal in magnitude at $\theta=\qty{45}{\degree}$ and \qty{135}{\degree}, and changes sign as the polarization crosses the plane perpendicular to the pump axis. This reversal follows directly from Eq.~\eqref{eq:anisotropic_optical_pumping_source}: the anisotropic correction is locked to the instantaneous polarization through $(\vb{s}\cdot\hat{\vb{n}})\hat{\vb{n}}$, so beyond $\theta=\qty{90}{\degree}$ it changes from pumping along the polarization to reducing it. Figure~\ref{fig:optical_pumping_drive_maps}(a) shows precisely this sign change and its growth with $P$, while panel (b) shows that $G_z$ remains positive and is largest near collinear pumping.

The source term alone, however, does not give the total difference between the anisotropic and scalar pumping dynamics. The repumping loss remains antiparallel to $\vb{P}$, but relative to the dynamic scalar model its magnitude is rescaled by replacing $q_\perp(P)$ with $q_\parallel(P)$. At the same instantaneous polarization, the dynamic scalar model gives $\left.\dv{}{t}P\right|_{\mathrm{OP}}^{(\mathrm{sc})}=R_{\mathrm{OP}}(\vb{s}\cdot\hat{\vb{n}}-P)/q_\perp(P)$. Subtracting this amplitude rate from the anisotropic result defines the total correction, for which we use the shorthand $\Delta\dot{P}_{\mathrm{OP}}$:
\begin{equation}
\begin{split}
    \Delta\dot{P}_{\mathrm{OP}}
    &\equiv\left.\dv{P}{t}\right|_{\mathrm{OP}}^{(\mathrm{an})}
    -\left.\dv{P}{t}\right|_{\mathrm{OP}}^{(\mathrm{sc})}\\
    &=R_{\mathrm{OP}}\left[\frac{1}{q_\parallel(P)}-\frac{1}{q_\perp(P)}\right]
    (\vb{s}\cdot\hat{\vb{n}}-P).
\end{split}
    \label{eq:optical_pumping_amplitude_correction}
\end{equation}
The corresponding vector correction is $\Delta\dot{\vb{P}}_{\mathrm{OP}}=\Delta\dot{P}_{\mathrm{OP}}\hat{\vb{n}}$: it acts along the instantaneous polarization, changing its amplitude without introducing a direct angular or phase term. Precession angle changes can arise indirectly through the resulting evolution of $P$. For circular pumping, the source correction reverses at $\theta=\pi/2$, but the total correction changes sign at $\cos\theta=P$. Below this boundary it enhances polarization buildup; above it, the combined source and repumping corrections enhance depolarization.

Figure~\ref{fig:optical_pumping_drive_maps}(c) shows $\Delta\dot{P}_{\mathrm{OP}}$ given by  Eq.~\eqref{eq:optical_pumping_amplitude_correction}. For large polarization tilted against the pump, its magnitude can exceed the scalar source magnitude $G_{\mathrm{sc}}=R_{\mathrm{OP}}/q_\perp(P)$, with equality marked by the white contour. The complete longitudinal pumping rate is rescaled by $q_\perp(P)/q_\parallel(P)$, reaching a factor of two at full polarization.

An interesting regime arises when a harmonic magnetic field transverse to the pump imposes a leading-order harmonic evolution of the angle of the polarization. We take $P^{(0)}(t)$ to be the approximate solution of the scalar Bloch equation and $P^{(1)}(t)\ll P^{(0)}(t)$ to be its anisotropy-induced amplitude correction. Because the correction in Eq.~\eqref{eq:optical_pumping_amplitude_correction} is longitudinal, both contributions share the same leading direction $\hat{\vb{n}}(t)$, and we write $\vb{P}(t)=[P^{(0)}(t)+P^{(1)}(t)]\hat{\vb{n}}(t)$. The magnetic drive sets $\hat{\vb{n}}(t)=\cos\!\big(\theta(t)\big)\hat{\vb{z}}+\sin\!\big(\theta(t)\big)\hat{\vb{x}}$, where $\theta(t)=\Theta\cos(\omega t)$, $\Theta$ is the angular oscillation amplitude, and $\omega$ is the drive frequency. The corresponding pump--polarization overlap is
\begin{equation}
\begin{split}
    \vb{s}\cdot\hat{\vb{n}}(t)
    &=\cos[\theta(t)]
    =\cos\!\left[\Theta\cos(\omega t)\right]\\
    &=J_0(\Theta)
    +2\sum_{k=1}^{\infty}(-1)^kJ_{2k}(\Theta)\cos(2k\omega t).
\end{split}
    \label{eq:pump_overlap_harmonics}
\end{equation}
The anisotropy-induced rate $\Delta\dot{P}_{\mathrm{OP}}$ drives the first-order amplitude correction $P^{(1)}$. Evaluating Eq.~\eqref{eq:optical_pumping_amplitude_correction} on the dominant trajectory and applying the Jacobi--Anger expansion gives
\begin{equation}
\begin{split}
    \Delta\dot{P}_{\mathrm{OP}}(t&)
    =R_{\mathrm{OP}}
    \left[
    \frac{1}{q_\parallel\big(P^{(0)}\big)}
    -\frac{1}{q_\perp\big(P^{(0)}\big)}
    \right]\\
    &\times\Bigg[
    A(t) +2\sum_{k=1}^{\infty}(-1)^kJ_{2k}(\Theta)
    \cos(2k\omega t)\Bigg],
\end{split}
    \label{eq:pumping_correction_harmonics}
\end{equation}
where $A(t)=J_0(\Theta)-P^{(0)}(t)$ and $J_k$ is the Bessel function of the first kind. The Jacobi--Anger terms contain a dc component and all even harmonics of the drive frequency. The even-harmonic content of the pump–polarization overlap is geometric and is not itself generated by the slowing-down anisotropy. Rather, the anisotropy converts this overlap into an additional longitudinal amplitude response through the prefactor in Eq.~\eqref{eq:pumping_correction_harmonics}, which vanishes in the scalar limit $q_\parallel=q_\perp$. The evolution of $P^{(0)}(t)$ and the polarization dependence of the slowing-down factors further modulate this response and introduce additional frequency mixing.

\section{Discussion and conclusion}
\label{sec:discussion}

We have shown that the commonly used scalar slowing-down description can produce a systematic error whenever the polarization magnitude changes appreciably. The origin is the distinction between the algebraic spin-temperature relation, which is scalar, and its differential response, which separates into the longitudinal and transverse factors $q_\parallel(P)$ and $q_\perp(P)$. During pump-free evolution, an incorrect longitudinal response changes the trajectory $P(t)$ and thereby modifies the accumulated precession angle and the inferred magnetic field. For the $^{87}\mathrm{Rb}$ conditions considered here, the dynamic scalar model reaches a magnetic field bias of about $4.2\%$, while the fixed scalar approximation produces an error of about $10.1\%$. Including the nuclear gyromagnetic correction reduces the residual anisotropic-model bias below $10^{-3}$ at the reference temperature. The residual decreases further at higher temperature, consistent with the spin-temperature approximation becoming more accurate as spin exchange becomes faster relative to the other dynamics.

The consequences extend beyond free polarization decay. Under continuous pumping, slowing-down anisotropy modifies the longitudinal pumping dynamics through the nonlinear correction identified in Eq.~\eqref{eq:optical_pumping_amplitude_correction}, which acts along the instantaneous polarization and therefore changes its amplitude without introducing an additional transverse torque. Under periodic driving, the pump–polarization geometry can already produce harmonic structure in the overlap $\mathbf s\cdot\hat{\mathbf n}(t)$. The slowing-down anisotropy converts this overlap into an additional longitudinal amplitude response that vanishes in the scalar limit. The polarization dependence of the two slowing-down factors further modulates this response and introduces additional frequency mixing.

These results motivate retaining separate $q_\parallel(P)$ and $q_\perp(P)$ in transient and modulated SERF magnetometry whenever the polarization enters the strongly polarized regime. The anisotropic Bloch equation preserves the simplicity of a three-component model while substantially reducing the error relative to the master equation. It also captures an anisotropy-induced contribution to the harmonic response that is absent from scalar slowing-down descriptions.

\medskip
\noindent\textit{Note added.---}After making these calculations we became aware of a preprint by Koutrouli, Vasilakis, and Mouloudakis  with some overlapping results \cite{Koutrouli2026}.  Their work focuses on the microscopic origin of the longitudinal and transverse slowing-down factors, and the present work addresses their consequences for finite-amplitude transient dynamics, magnetic-field inference, and driven optical pumping.

\begin{acknowledgments}
Work supported by European Commission projects Field-SEER (ERC 101097313) and QUANTIFY (101135931); Horizon Europe (Q-Planet, 101291743), Chips Joint Undertaking (Chips JU) and the Spanish Ministry for Digital Transformation and Civil Service under reference CJU-010200-2026-6, as part of the Recovery, Transformation and Resilience Plan (PRTR),  Spanish Ministry of Science MCIN project SALVIA (PID2024-158479NB-I00),  ``Severo Ochoa'' Center of Excellence CEX2024-001490-S [MICIU/AEI/10.13039/501100011033];  Generalitat de Catalunya through the CERCA program,  DURSI grant No. 2021 SGR 01453 and QSENSE (GOV/51/2022).  Fundaci\'{o} Privada Cellex; Fundaci\'{o} Mir-Puig. The project that gave rise to these results received the support of a fellowship from the “la Caixa” Foundation (ID 100010434). The fellowship code is LCF/BQ/DI24/12070013. OpenAI ChatGPT (GPT-5.6) was used during manuscript preparation to assist with language editing and the clarification of scientific arguments. GPT-5.6 Sol was additionally used to assist with debugging numerical code. All text and code modifications were critically reviewed and independently verified by the authors, who take full responsibility for the final manuscript.
\end{acknowledgments}

\section*{Data Availability Statement}

The data generated by the numerical simulations in this study are not publicly available but are available from the corresponding author upon reasonable request.


\appendix
\section{Derivation of the anisotropic Bloch equation}
\label{app:bloch_derivation}

This appendix derives Eq.~\eqref{eq:anisotropic_bloch} directly from the
master equation, keeping only the steps needed to identify the physical
closure and the origin of the two slowing-down factors.
For completeness, we restate the standard ground-state master equation
from which the derivation starts \cite{Appelt1998}:
\begin{equation}
\begin{split}
    \dv{\rho}{t}
    &={}
    -i\comm{\widehat{\mathcal{H}}(t)}{\rho}
    +R_{\mathrm{SD}}\left[\varphi(\rho)-\rho\right]
    \\
    &\quad+
    R_{\mathrm{SE}}
    \left[
    \varphi(\rho)
    \left(
    \widehat{\mathds{1}}+4\expval{\widehat{\vb{S}}}\cdot\widehat{\vb{S}}
    \right)
    -\rho
    \right]
    \\
    &\quad+
    R_{\mathrm{OP}}
    \left[
    \varphi(\rho)
    \left(
    \widehat{\mathds{1}}+2\vb{s}\cdot\widehat{\vb{S}}
    \right)
    -\rho
    \right].
    \label{eq:master_serf_appendix}
\end{split}
\end{equation}

\subsection{First moments of the master equation}

For a spin-$1/2$ electron, the map appearing in
Eq.~\eqref{eq:master_serf_appendix} can be written as
\begin{equation}
\varphi(\rho)
=
\frac{\widehat{\mathds{1}}_S}{2}\otimes\rho_I,
\qquad
\rho_I=\operatorname{Tr}_S\rho .
\label{eq:phi_reduced}
\end{equation}
It therefore removes electronic orientation while preserving the reduced
nuclear state. The identities needed below follow immediately:
\begin{equation}
\operatorname{Tr}[\widehat{S}_a\varphi(\rho)]=0,
\qquad
\operatorname{Tr}[\widehat{S}_a\varphi(\rho)\widehat{S}_b]=\frac{\delta_{ab}}{4},
\label{eq:phi_S_identities}
\end{equation}
and
\begin{equation}
\operatorname{Tr}[\widehat{I}_a\varphi(\rho)]=\expval{\widehat{I}_a},
\qquad
\operatorname{Tr}[\widehat{I}_a\varphi(\rho)\widehat{S}_b]=0.
\label{eq:phi_I_identities}
\end{equation}

Using $\dv{}{t}\expval{\widehat{S}_a}=\operatorname{Tr}(\widehat{S}_a\dv{}{t}\rho)$ and the
angular-momentum commutation relations, the hyperfine Hamiltonian gives
\begin{equation}
\left.\dv{}{t}\expval{\widehat{\vb{S}}}\right|_{\mathrm{hf}}
=
\omega_{\mathrm{hf}}\expval{\widehat{\vb{I}}\times\widehat{\vb{S}}} ,
\end{equation}
while the electronic Zeeman term gives
\begin{equation}
\left.\dv{}{t}\expval{\widehat{\vb{S}}}\right|_{B,e}
=
\gamma_e\,\vb{B}\times\expval{\widehat{\vb{S}}} .
\end{equation}
The nuclear Zeeman Hamiltonian acts only on the nuclear subspace and has
no direct contribution to $\dv{}{t}\expval{\widehat{\vb{S}}}$.

Equations~\eqref{eq:phi_S_identities} give the dissipative first moments.
Spin destruction relaxes the electronic orientation,
\begin{equation}
\left.\dv{}{t}\expval{\widehat{\vb{S}}}\right|_{\mathrm{SD}}
=
-R_{\mathrm{SD}}\expval{\widehat{\vb{S}}} ,
\end{equation}
whereas the spin-exchange loss and repopulation terms cancel in the mean
electron spin,
\begin{equation}
\left.\dv{}{t}\expval{\widehat{\vb{S}}}\right|_{\mathrm{SE}}=0.
\end{equation}
Although spin exchange does not directly change the electronic first moment, it rapidly redistributes population among the remaining spin degrees of freedom and is responsible for the spin-temperature closure used below. Optical pumping gives
\begin{equation}
\left.\dv{}{t}\expval{\widehat{\vb{S}}}\right|_{\mathrm{OP}}
=
R_{\mathrm{OP}}
\left(\frac{\vb{s}}{2}-\expval{\widehat{\vb{S}}}\right).
\end{equation}
The exact electronic first-moment equation is therefore
\begin{equation}
\begin{aligned}
\dv{}{t}\expval{\widehat{\vb{S}}}
={}&\omega_{\mathrm{hf}}\expval{\widehat{\vb{I}}\times\widehat{\vb{S}}}+\gamma_e\vb{B}\times\expval{\widehat{\vb{S}}}\\
&-R_{\mathrm{SD}}\expval{\widehat{\vb{S}}}+R_{\mathrm{OP}}
\left(\frac{\vb{s}}{2}-\expval{\widehat{\vb{S}}}\right).
\end{aligned}
\label{eq:S_exact_appendix}
\end{equation}
The correlation $\expval{\widehat{\vb{I}}\times\widehat{\vb{S}}}$ prevents closure in
terms of the electron polarization alone.

The same calculation for the nuclear first moment gives
\begin{equation}
\dv{}{t}\expval{\widehat{\vb{I}}}
=
-\omega_{\mathrm{hf}}\expval{\widehat{\vb{I}}\times\widehat{\vb{S}}}
+\gamma_I\vb{B}\times\expval{\widehat{\vb{I}}} .
\label{eq:I_exact_appendix}
\end{equation}
The dissipative terms in Eq.~\eqref{eq:master_serf_appendix} do not directly change
$\expval{\widehat{\vb{I}}}$ for the collision model used here. Adding
Eqs.~\eqref{eq:S_exact_appendix} and \eqref{eq:I_exact_appendix}
eliminates the internal hyperfine torque. With $\widehat{\vb{F}}=\widehat{\vb{I}}+\widehat{\vb{S}}$ one obtains Eq.~\eqref{eq:F_exact}.
\begin{equation}
\begin{split}
\dv{}{t}\expval{\widehat{\vb{F}}}
&=
\gamma_e\vb{B}(t)\times\expval{\widehat{\vb{S}}}
+\gamma_I\vb{B}(t)\times\expval{\widehat{\vb{I}}}\\
&\quad-R_{\mathrm{SD}}\expval{\widehat{\vb{S}}}
+R_{\mathrm{OP}}\left(\frac{\vb{s}}{2}-\expval{\widehat{\vb{S}}}\right).
\end{split}
\end{equation}
This equation is exact at the level of
first moments; the approximation enters only in spin-temperature closure, when
$\expval{\widehat{\vb{I}}}$ and $\expval{\widehat{\vb{F}}}$ are expressed in terms
of $\vb{P}$.

\subsection{Spin-temperature closure}

Rapid spin exchange in the SERF regime constrains the density matrix
close to
\begin{equation}
\rho_{\mathrm{ST}}
=
\frac{\exp(\beta\hat{\vb{n}}\cdot\widehat{\vb{F}})}{Z},
\label{eq:rho_ST_appendix}
\end{equation}
where $Z$ is the partition function. Because nuclear and electronic spin operators commute,
\begin{equation}
\exp[\beta\hat{\vb{n}}\cdot(\widehat{\vb{I}}+\widehat{\vb{S}})]
=
\exp(\beta\hat{\vb{n}}\cdot\widehat{\vb{I}})
\exp(\beta\hat{\vb{n}}\cdot\widehat{\vb{S}}),
\label{eq:spin_temperature_factorization}
\end{equation}
so all first-rank spin moments are collinear with $\hat{\vb{n}}$.
We define the electron polarization as
$\vb{P}\equiv\expval{\widehat{\vb{S}}}/S=2\expval{\widehat{\vb{S}}}$.
For $S=1/2$, evaluating the two spin-temperature weights associated with
the eigenvalues $m_S=\pm1/2$ in Eq.~\eqref{eq:spin_temperature_factorization}
gives the first relation below, while its inversion gives the second:
\begin{equation}
P=\tanh\!\left(\frac{\beta}{2}\right),
\qquad
\beta=2\,\operatorname{arctanh}(P).
\label{eq:beta_P}
\end{equation}

For an arbitrary nuclear spin $I$, the nuclear factor in
Eq.~\eqref{eq:spin_temperature_factorization} has the partition function
\begin{equation}
Z_I(\beta)=\sum_{m=-I}^{I}\exp(\beta m).
\label{eq:general_nuclear_partition}
\end{equation}
Because the nuclear first moment is parallel to $\hat{\vb{n}}$, its
magnitude follows from the logarithmic derivative of the partition function,
\begin{equation}
\expval{\widehat{\vb{I}}}\cdot\hat{\vb{n}}
=\pdv{}{\beta}\ln Z_I(\beta).
\label{eq:general_nuclear_moment}
\end{equation}

In particular, $^{87}\mathrm{Rb}$ has $I=3/2$, for which
\begin{equation}
Z_{3/2}(\beta)
=
2\cosh\!\left(\frac{3\beta}{2}\right)
+
2\cosh\!\left(\frac{\beta}{2}\right),
\end{equation}
and substituting this result into Eq.~\eqref{eq:general_nuclear_moment}
and eliminating $\beta$ with Eq.~\eqref{eq:beta_P} gives
\begin{equation}
\expval{\widehat{\vb{I}}}
=
\frac{5+P^2}{2(1+P^2)}\,\vb{P}.
\label{eq:I_P_appendix}
\end{equation}
Adding $\expval{\widehat{\vb{S}}}=\vb{P}/2$ yields
\begin{equation}
\expval{\widehat{\vb{F}}}
=
\frac{3+P^2}{1+P^2}\,\vb{P}
=
F(P)\hat{\vb{n}}.
\label{eq:F_P_appendix}
\end{equation}

\subsection{Differential response on the spin-temperature manifold}

The closure in Eq.~\eqref{eq:F_P_appendix} is nonlinear. Differentiating
$\vb{P}=P\hat{\vb{n}}$ and
$\expval{\widehat{\vb{F}}}=F(P)\hat{\vb{n}}$ gives
\begin{align}
\dv{\vb{P}}{t}
&=
\dv{P}{t}\,\hat{\vb{n}}
+
P\,\dv{\hat{\vb{n}}}{t},
\\
\dv{}{t}\expval{\widehat{\vb{F}}}
&=
\dv{F}{P}\dv{P}{t}\,\hat{\vb{n}}
+
F(P)\dv{\hat{\vb{n}}}{t} .
\end{align}
The first terms are parallel to $\vb{P}$ and the second are transverse.
Hence
\begin{equation}
\left[\dv{}{t}\expval{\widehat{\vb{F}}}\right]_{\parallel}
=
\dv{F}{P}\left[\dv{\vb{P}}{t}\right]_{\parallel},
\label{eq:parallel_response_appendix}
\end{equation}
and
\begin{equation}
\left[\dv{}{t}\expval{\widehat{\vb{F}}}\right]_{\perp}
=
\frac{F(P)}{P}\left[\dv{\vb{P}}{t}\right]_{\perp}.
\label{eq:transverse_response_appendix}
\end{equation}
We identify the two slowing-down factors as follows; for
$^{87}\mathrm{Rb}$, where $I=3/2$, Eq.~\eqref{eq:F_P_appendix} gives
\begin{align}
q_\perp(P)=2\frac{F(P)}{P}
&\xrightarrow{\ I=3/2\ }
2\frac{3+P^2}{1+P^2},
\label{eq:q_perp_appendix}
\\
q_\parallel(P)=2\dv{F}{P}
&\xrightarrow{\ I=3/2\ }
2\frac{3+P^4}{(1+P^2)^2}.
\label{eq:q_parallel_appendix}
\end{align}
The corresponding slowing-down factors for the other relevant nuclear
spins are collected in Table~\ref{tab:slowing_factors}.

Finally, substituting Eq.~\eqref{eq:I_P_appendix} together with
$\expval{\widehat{\vb{S}}}=\vb{P}/2$ into Eq.~\eqref{eq:F_exact}
expresses its right-hand side entirely in terms of $\vb{P}$. We then
split $\dv{}{t}\vb{P}$ into its transverse and longitudinal components
using Eqs.~\eqref{eq:parallel_response_appendix} and
\eqref{eq:transverse_response_appendix}:
\begin{equation}
\begin{aligned}
\dv{\vb{P}}{t}
={}&
\frac{1}{q_\perp(P)}
\left[2\dv{}{t}\expval{\widehat{\vb{F}}}\right]_{\perp}
+
\frac{1}{q_\parallel(P)}
\left[2\dv{}{t}\expval{\widehat{\vb{F}}}\right]_{\parallel}
\\
={}&
\frac{1}{q_\perp(P)}
\left\{
\left[
\gamma_e+\gamma_I\frac{5+P^2}{1+P^2}
\right]\vb{B}(t)\times\vb{P}
+R_{\mathrm{OP}}\vb{s}_\perp
\right\}
\\
&+
\frac{1}{q_\parallel(P)}
\left[
-R_{\mathrm{SD}}\vb{P}
+R_{\mathrm{OP}}(\vb{s}_\parallel-\vb{P})
\right],
\end{aligned}
\label{eq:anisotropic_bloch_appendix}
\end{equation}
which is Eq.~\eqref{eq:anisotropic_bloch}.

\subsection{Different alkali-metal species}
\begin{figure}[H]
    \centering
    \includegraphics[width=\columnwidth]{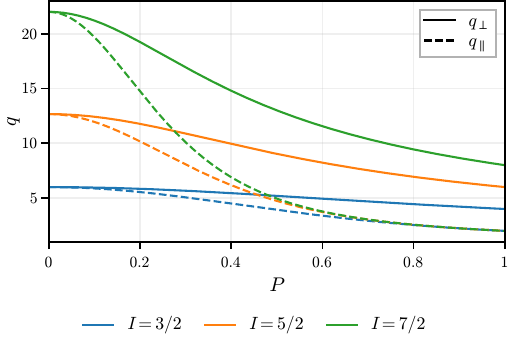}
    \caption{Nuclear slowing-down factors versus electron polarization for $I=3/2$, $5/2$, and $7/2$. Solid curves show $q_\perp$ and dashed curves show $q_\parallel$; color identifies the nuclear spin. Exact expressions are collected in Table~\ref{tab:slowing_factors}.}
    \label{fig:slowing_factors}
\end{figure}

The nuclear spin entering the slowing-down factors is isotope dependent.
The cases shown in Fig.~\ref{fig:slowing_factors} and summarized in
Table~\ref{tab:slowing_factors} cover the principal isotopes used in
alkali-vapor experiments: $^{7}\mathrm{Li}$, $^{23}\mathrm{Na}$,
$^{39}\mathrm{K}$, $^{41}\mathrm{K}$, and $^{87}\mathrm{Rb}$ have
$I=3/2$; $^{85}\mathrm{Rb}$ has $I=5/2$; and $^{133}\mathrm{Cs}$ has
$I=7/2$. The other naturally occurring isotopes of lithium and potassium,
$^{6}\mathrm{Li}$ and $^{40}\mathrm{K}$, have $I=1$ and $I=4$,
respectively, and are not included among the three cases plotted.

\begin{table*}[t]
    \centering
    \caption{Transverse and longitudinal nuclear slowing-down factors on the spin-temperature manifold. The $I=3/2$ expressions also apply to $^{7}\mathrm{Li}$, $^{23}\mathrm{Na}$, $^{39}\mathrm{K}$, and $^{41}\mathrm{K}$.}
    \label{tab:slowing_factors}
    \renewcommand{\arraystretch}{1.8}
    \begin{tabular}{clcc}
        \toprule
        $I$ & Representative isotope & $q_\perp(P)$ & $q_\parallel(P)$ \\
        \midrule
        $3/2$ & $^{87}\mathrm{Rb}$
        & $\displaystyle 2\frac{P^2+3}{P^2+1}$
        & $\displaystyle 2\frac{P^4+3}{(P^2+1)^2}$ \\[1ex]
        $5/2$ & $^{85}\mathrm{Rb}$
        & $\displaystyle 2\frac{3P^4+26P^2+19}{(P^2+3)(3P^2+1)}$
        & $\displaystyle 2\frac{9P^8+12P^6+134P^4+44P^2+57}{(P^2+3)^2(3P^2+1)^2}$ \\[1ex]
        $7/2$ & $^{133}\mathrm{Cs}$
        & $\displaystyle 2\frac{P^6+17P^4+35P^2+11}{(P^2+1)(P^4+6P^2+1)}$
        & $\displaystyle 2\frac{P^{12}+4P^{10}+49P^8+64P^6+99P^4+28P^2+11}{(P^2+1)^2(P^4+6P^2+1)^2}$ \\
        \bottomrule
    \end{tabular}
\end{table*}

\section{Amplitude--precession decomposition and longitudinal decay}
\label{app:derivation}
\label{app:fid_derivation}

\subsection{Amplitude and precession angle in a transverse field}

With $R_{\mathrm{OP}}=0$ and $\vb{B}=B\hat{\vb{x}}$, Eq.~\eqref{eq:anisotropic_bloch} becomes
\begin{equation}
    \dv{\vb{P}}{t}
    =\Omega(P)\hat{\vb{x}}\times\vb{P}
    -\frac{R_{\mathrm{SD}}}{q_\parallel(P)}\vb{P},
    \label{eq:free_bloch_appendix}
\end{equation}
where the full precession coefficient is
\begin{equation}
    \Omega(P)=\frac{B}{q_\perp(P)}
    \left[\gamma_e+\gamma_I\frac{5+P^2}{1+P^2}\right].
    \label{eq:fid_frequency_appendix}
\end{equation}
The $x$ component satisfies $\dv{}{t}P_x=-R_{\mathrm{SD}}P_x/q_\parallel(P)$, so $P_x(0)=0$ implies $P_x(t)=0$. The $yz$ plane is therefore invariant. In addition to $\hat{\vb{u}}(\phi)$ from Eq.~\eqref{eq:fid_trajectory}, introduce its angular tangent,
\begin{equation}
    \hat{\vb{v}}(\phi)
    =-\sin\phi\,\hat{\vb{z}}-\cos\phi\,\hat{\vb{y}}.
\end{equation}
These vectors are orthonormal and satisfy $\dv{}{\phi}\hat{\vb{u}}=\hat{\vb{v}}$ and $\hat{\vb{x}}\times\hat{\vb{u}}=\hat{\vb{v}}$. Differentiating $\vb{P}=P\hat{\vb{u}}$ and substituting it into Eq.~\eqref{eq:free_bloch_appendix} gives
\begin{equation}
    \dv{P}{t}\hat{\vb{u}}+P\dv{\phi}{t}\hat{\vb{v}}
    =-\frac{R_{\mathrm{SD}}}{q_\parallel(P)}P\hat{\vb{u}}
    +\Omega(P)P\hat{\vb{v}}.
\end{equation}
Projection onto the radial and angular basis vectors yields $\dv{}{t}P=-R_{\mathrm{SD}}P/q_\parallel(P)$ and $\dv{}{t}\phi=\Omega(P)$ for $P>0$. Neglecting $\gamma_I$ recovers Eqs.~\eqref{eq:fid_amplitude} and \eqref{eq:fid_phase}. Including the nuclear Zeeman term changes only the precession-angle equation. The accumulated precession can be obtained by continuously unwrapping $\arg[P_z(t)-iP_y(t)]$, consistent with $P_z-iP_y=P e^{i\phi}$.

\section{Simulation parameters}
\label{app:common_simulation_parameters}
\label{app:numerics}

The conclusions of this work apply directly to any alkali-metal species operated in the SERF regime. Owing to the widespread use of rubidium in SERF magnetometry, however, we present the numerical results for $^{87}\mathrm{Rb}$ and use the vapor-cell parameters reported in Ref.~\cite{Raghavan2024}. Table~\ref{tab:common_simulation_parameters} collects the parameters held fixed across the simulations. Quantities varied in individual calculations, including the applied and modulated magnetic fields, pump power, initial polarization, and evolution time, are omitted from the table and specified separately below.

\begin{table}[H]
    \centering
    \caption{Atomic and vapor-cell parameters for the $\mathrm{N}_2$-buffered cell \cite{Raghavan2024} shared by the numerical simulations, together with selected rates calculated from them.}
    \label{tab:common_simulation_parameters}
    \renewcommand{\arraystretch}{1.15}
    \begin{tabular}{lcc}
        \toprule
        Parameter & Symbol & Value \\
        \midrule
        Cell temperature & $T$ & \qty{190}{\degreeCelsius}$^{*}$ \\
        Filling pressure & $P_{\mathrm{fill}}$ & \qty{2250.19}{\torr} \\
        Filling temperature & $T_{\mathrm{fill}}$ & \qty{300.15}{\kelvin} \\
        Pump-beam diameter & $d$ & \qty{2}{\milli\metre} \\
        \midrule
        \multirow{3}{*}{Spin-exchange rate}
            & $R_{\mathrm{SE}}(\qty{130}{\degreeCelsius})$ & \qty{2.997e4}{\per\second} \\
            & $R_{\mathrm{SE}}(\qty{190}{\degreeCelsius})$ & \qty{5.555e5}{\per\second} \\
            & $R_{\mathrm{SE}}(\qty{230}{\degreeCelsius})$ & \qty{2.631e6}{\per\second} \\
        \midrule
        \multirow{3}{*}{Spin-destruction rate}
            & $R_{\mathrm{SD}}(\qty{130}{\degreeCelsius})$ & \qty{485.3}{\per\second} \\
            & $R_{\mathrm{SD}}(\qty{190}{\degreeCelsius})$ & \qty{960.9}{\per\second} \\
            & $R_{\mathrm{SD}}(\qty{230}{\degreeCelsius})$ & \qty{2730.0}{\per\second} \\
        \midrule
        Optical-pumping rate
            & $R_{\mathrm{OP}}(\qty{1}{\micro\watt})$ & \qty{33.51}{\per\second}$^{\dagger}$ \\
        \bottomrule
    \end{tabular}
    \vspace{1pt}

    \begin{minipage}{\columnwidth}
        \footnotesize $^{*}$The temperature scan in Fig.~\ref{fig:fid_phase_bias_and_fisher_vs_initial_polarization}(c) instead uses the values specified below.

        $^{\dagger}$For other pump powers, $R_{\mathrm{OP}}$ scales linearly with the optical power and can be obtained directly from the tabulated value.
    \end{minipage}
\end{table}

\subsection{Figure~\ref{fig:relaxation_comparison}}
The two initial states were prepared by evolving the full master equation for \qty{100}{\milli\second} from the unpolarized state, in zero magnetic field and under circularly polarized pumping along $+\hat{\vb{z}}$. Pump powers of \qty{12}{\micro\watt} and \qty{10}{\milli\watt} produced $P_0\simeq0.295$ and $0.997$, respectively. Denoting the resulting density matrix by $\rho_0$, the master-equation decay was initialized directly with $\rho_0$, whereas each reduced model was initialized with its electronic polarization,
\begin{equation}
    \vb{P}(0)=2\operatorname{Tr}\!\left(\rho_0\widehat{\vb{S}}\right).
    \label{eq:prepared_initial_polarization}
\end{equation}
At $t=0$, the pump was switched off and a static transverse field $\vb{B}=B\hat{\vb{x}}$, with $B=\qty{0}{\nano\tesla}$ or \qty{100}{\nano\tesla}, was applied for the \qty{15}{\milli\second} free-polarization-decay interval.

\subsection{Figure~\ref{fig:fid_phase_bias_and_fisher_vs_initial_polarization}}
Each value of $P_0$ was generated by the same master-equation preparation used for Fig.~\ref{fig:relaxation_comparison}, now applied for \qty{20}{\milli\second} at zero field. The pump powers were \numlist{0.5;2;4;7;10.5;15;21;27;34;40;53;65;80;100;140;200;350;700}\,\si{\micro\watt} and \qty{10}{\milli\watt}. The prepared density matrix and the polarization defined in Eq.~\eqref{eq:prepared_initial_polarization} were then used to initialize the master equation and the reduced models, respectively. After preparation, the pump was switched off and a static field $\vb{B}=\qty{5}{\nano\tesla}\hat{\vb{x}}$ was applied. Panels (a) and (b) use $T=\qty{190}{\degreeCelsius}$; panel (c) repeats the preparation and decay at $T=\qtylist{130;140;160;190;230}{\degreeCelsius}$.

For each prepared state, the analysis interval ends when the magnitude of the master-equation polarization first reaches $P_0 e^{-2}$. This cutoff avoids accumulating precession from the late-time part of the transient, where an experimental signal would approach the noise floor and cease to provide reliable precession-angle information.

\subsection{Fisher information in Fig.~\ref{fig:fid_phase_bias_and_fisher_vs_initial_polarization}(b)}
The Fisher information is evaluated entirely from master-equation trajectories over the same cutoff interval. We assume additive Gaussian polarization noise with a common variance $\sigma_P^2$, independent between polarization components and measurement times. For this measurement model,
\begin{equation}
    \mathcal{F}_{B}
    =\frac{1}{\sigma_P^2}
    \sum_{t_k\leq t_{\mathrm{cut}}}\sum_{a\in\{x,y,z\}}
    \left[
        \frac{\partial P_a(t_k;B)}{\partial B}
    \right]^2.
    \label{eq:appendix_fisher_information}
\end{equation}
The derivatives are obtained by a central difference of trajectories at $B\pm\delta B$, with $B=\qty{5}{\nano\tesla}$ and $\delta B=\qty{0.01}{\nano\tesla}$. Both trajectories start from the same density matrix prepared at zero field, so the finite difference probes only the response of the subsequent pump-free evolution to the applied field. Because the simulations do not specify an absolute noise level, we set $\sigma_P=1$ and plot the information normalized over the prepared-polarization scan,
\begin{equation}
    \widetilde{\mathcal{F}}_{B}(P_0)
    =\frac{\mathcal{F}_{B}(P_0)}
    {\max_{P_0}\mathcal{F}_{B}(P_0)}.
    \label{eq:appendix_normalized_fisher_information}
\end{equation}
This normalization removes the arbitrary common noise scale; consequently, panel (b) reports Fisher information in arbitrary units.

\bibliography{references}

@article{Mouloudakis2024,
  title = {Anomalous noise spectra in a spin-exchange-relaxation-free alkali-metal vapor},
  author = {Mouloudakis, K. and Kong, J. and Sierant, A. and Arkin, E. and Hern\'andez Ruiz, M. and Jim\'enez-Mart\'{\i}nez, R. and Mitchell, M. W.},
  journal = {Phys. Rev. A},
  volume = {109},
  issue = {4},
  pages = {L040802},
  numpages = {5},
  year = {2024},
  month = {Apr},
  publisher = {American Physical Society},
  doi = {10.1103/PhysRevA.109.L040802},
  url = {https://link.aps.org/doi/10.1103/PhysRevA.109.L040802}
}

@article{Limes2018,
  title = {$^{3}\mathrm{He}\text{\ensuremath{-}}^{129}\mathrm{Xe}$ Comagnetometery using $^{87}\mathrm{Rb}$ Detection and Decoupling},
  author = {Limes, M. E. and Sheng, D. and Romalis, M. V.},
  journal = {Phys. Rev. Lett.},
  volume = {120},
  issue = {3},
  pages = {033401},
  numpages = {5},
  year = {2018},
  month = {Jan},
  publisher = {American Physical Society},
  doi = {10.1103/PhysRevLett.120.033401},
  url = {https://link.aps.org/doi/10.1103/PhysRevLett.120.033401}
}

@article{Wei2023,
  title = {Ultrasensitive Atomic Comagnetometer with Enhanced Nuclear Spin Coherence},
  author = {Wei, Kai and Zhao, Tian and Fang, Xiujie and Xu, Zitong and Liu, Chang and Cao, Qian and Wickenbrock, Arne and Hu, Yanhui and Ji, Wei and Fang, Jiancheng and Budker, Dmitry},
  journal = {Phys. Rev. Lett.},
  volume = {130},
  issue = {6},
  pages = {063201},
  numpages = {6},
  year = {2023},
  month = {Feb},
  publisher = {American Physical Society},
  doi = {10.1103/PhysRevLett.130.063201},
  url = {https://link.aps.org/doi/10.1103/PhysRevLett.130.063201}
}

@article{Bodenstedt2021,
  author  = {Bodenstedt, Sven and Mitchell, Morgan W. and Tayler, Michael C. D.},
  title   = {Fast-field-cycling ultralow-field nuclear magnetic relaxation dispersion},
  journal = {Nature Communications},
  volume  = {12},
  number  = {1},
  pages   = {4041},
  year    = {2021},
  doi     = {10.1038/s41467-021-24248-9}
}

@article{Afach2021,
  author  = {Afach, Samer and Buchler, Ben C. and Budker, Dmitry and Dailey, Conner and Derevianko, Andrei and Dumont, Vincent and Figueroa, Nataniel L. and Gerhardt, Ilja and Gruji{\'c}, Zoran D. and Guo, Hong and Hao, Chuanpeng and Hamilton, Paul S. and Hedges, Morgan and Jackson Kimball, Derek F. and Kim, Dongok and Khamis, Sami and Kornack, Thomas and Lebedev, Victor and Lu, Zheng-Tian and Masia-Roig, Hector and Monroy, Madeline and Padniuk, Mikhail and Palm, Christopher A. and Park, Sun Yool and Paul, Karun V. and Penaflor, Alexander and Peng, Xiang and Pospelov, Maxim and Preston, Rayshaun and Pustelny, Szymon and Scholtes, Theo and Segura, Perrin C. and Semertzidis, Yannis K. and Sheng, Dong and Shin, Yun Chang and Smiga, Joseph A. and Stalnaker, Jason E. and Sulai, Ibrahim and Tandon, Dhruv and Wang, Tao and Weis, Antoine and Wickenbrock, Arne and Wilson, Tatum and Wu, Teng and Wurm, David and Xiao, Wei and Yang, Yucheng and Yu, Dongrui and Zhang, Jianwei},
  title   = {Search for topological defect dark matter with a global network of optical magnetometers},
  journal = {Nature Physics},
  volume  = {17},
  number  = {12},
  pages   = {1396--1401},
  year    = {2021},
  doi     = {10.1038/s41567-021-01393-y}
}

@article{Boto2018,
  title = {Moving magnetoencephalography towards real-world applications with a wearable system},
  author = {Boto, E. and Holmes, N. and Leggett, J. and Roberts, G. and Shah, V. and Meyer, S. S. and Duque Mu{\~n}oz, L. and Mullinger, K. J. and Tierney, T. M. and Bestmann, S. and Barnes, G. R. and Bowtell, R. and Brookes, M. J.},
  journal = {Nature},
  volume = {555},
  pages = {657--661},
  year = {2018},
  doi = {10.1038/nature26147}
}

@article{Tierney2019,
  title = {Optically pumped magnetometers: From quantum origins to multi-channel magnetoencephalography},
  author = {Tierney, T. M. and Holmes, N. and Mellor, S. and L{\'o}pez, J. D. and Roberts, G. and Hill, R. M. and Boto, E. and Leggett, J. and Shah, V. and Brookes, M. J. and Bowtell, R. and Barnes, G. R.},
  journal = {NeuroImage},
  volume = {199},
  pages = {598--608},
  year = {2019},
  doi = {10.1016/j.neuroimage.2019.05.063}
}

@article{Vasilakis2009,
  title = {Limits on new long range nuclear spin-dependent forces set with a {K}-{$^{3}$He} comagnetometer},
  author = {Vasilakis, G. and Brown, J. M. and Kornack, T. W. and Romalis, M. V.},
  journal = {Phys. Rev. Lett.},
  volume = {103},
  pages = {261801},
  year = {2009},
  doi = {10.1103/PhysRevLett.103.261801}
}

@article{Zhang2026,
  title = {{$^{3}$He}--{$^{21}$Ne} {Ramsey} Comagnetometer with {Sub-nHz} Frequency Resolution},
  author = {Zhang, S. and Wang, J. and Sun, G. and van de Wetering, J. J. and Romalis, M. V.},
  journal = {Phys. Rev. Lett.},
  volume = {136},
  pages = {203201},
  year = {2026},
  doi = {10.1103/wqqq-s2bz}
}

@article{Allred2002,
  title = {High-sensitivity atomic magnetometer unaffected by spin-exchange relaxation},
  author = {Allred, J. C. and Lyman, R. N. and Kornack, T. W. and Romalis, M. V.},
  journal = {Phys. Rev. Lett.},
  volume = {89},
  pages = {130801},
  year = {2002},
  doi = {10.1103/PhysRevLett.89.130801}
}

@article{Kominis2003,
  title = {A subfemtotesla multichannel atomic magnetometer},
  author = {Kominis, I. K. and Kornack, T. W. and Allred, J. C. and Romalis, M. V.},
  journal = {Nature},
  volume = {422},
  pages = {596--599},
  year = {2003},
  doi = {10.1038/nature01484}
}

@article{Savukov2005,
  title = {Effects of spin-exchange collisions in a high-density alkali-metal vapor in low magnetic fields},
  author = {Savukov, I. M. and Romalis, M. V.},
  journal = {Phys. Rev. A},
  volume = {71},
  pages = {023405},
  year = {2005},
  doi = {10.1103/PhysRevA.71.023405}
}

@article{Tang2021,
  title = {Transient dynamics of atomic spin in the spin-exchange-relaxation-free regime},
  author = {Tang, J. and Yin, Y. and Zhai, Y. and Zhou, B. and Han, B. and Yang, H. and Liu, G.},
  journal = {Opt. Express},
  volume = {29},
  pages = {8333},
  year = {2021},
  doi = {10.1364/OE.418776}
}

@article{Padniuk2022,
  title = {Response of atomic spin-based sensors to magnetic and nonmagnetic perturbations},
  author = {Padniuk, M. and Kopciuch, M. and Cipolletti, R. and Wickenbrock, A. and Budker, D. and Pustelny, S.},
  journal = {Sci. Rep.},
  volume = {12},
  number = {1},
  pages = {324},
  year = {2022},
  doi = {10.1038/s41598-021-03609-w},
  url = {https://doi.org/10.1038/s41598-021-03609-w}
}

@article{Happer1977,
  title = {Effect of rapid spin exchange on the magnetic-resonance spectrum of alkali vapors},
  author = {Happer, W. and Tam, A. C.},
  journal = {Phys. Rev. A},
  volume = {16},
  number = {5},
  pages = {1877--1891},
  year = {1977},
  doi = {10.1103/PhysRevA.16.1877},
  url = {https://link.aps.org/doi/10.1103/PhysRevA.16.1877}
}

@article{Appelt1998,
  title = {Theory of spin-exchange optical pumping of {$^{3}$He} and {$^{129}$Xe}},
  author = {Appelt, S. and Ben-Amar Baranga, A. and Erickson, C. J. and Romalis, M. V. and Young, A. R. and Happer, W.},
  journal = {Phys. Rev. A},
  volume = {58},
  number = {2},
  pages = {1412--1439},
  year = {1998},
  doi = {10.1103/PhysRevA.58.1412},
  url = {https://link.aps.org/doi/10.1103/PhysRevA.58.1412}
}

@article{Ledbetter2008,
  title = {Spin-exchange-relaxation-free magnetometry with {Cs} vapor},
  author = {Ledbetter, M. P. and Savukov, I. M. and Acosta, V. M. and Budker, D. and Romalis, M. V.},
  journal = {Phys. Rev. A},
  volume = {77},
  number = {3},
  pages = {033408},
  year = {2008},
  doi = {10.1103/PhysRevA.77.033408},
  url = {https://link.aps.org/doi/10.1103/PhysRevA.77.033408}
}

@article{Raghavan2024,
  title = {Functionalized millimeter-scale vapor cells for alkali-metal spectroscopy and magnetometry},
  author = {Raghavan, Harini and Tayler, Michael C. D. and Mouloudakis, Kostas and Rae, Rachel and L{\"a}hteenm{\"a}ki, Sami and Zetter, Rasmus and Laine, Petteri and Haesler, Jacques and Balet, Laurent and Overstolz, Thomas and Karlen, Sylvain and Mitchell, Morgan W.},
  journal = {Phys. Rev. Appl.},
  volume = {22},
  number = {4},
  pages = {044011},
  year = {2024},
  month = {Oct},
  publisher = {American Physical Society},
  doi = {10.1103/PhysRevApplied.22.044011},
  url = {https://link.aps.org/doi/10.1103/PhysRevApplied.22.044011}
}

@article{Koutrouli2026,
  title = {Nuclear slowing-down factors in alkali-metal vapors},
  author = {Koutrouli, Vasiliki and Vasilakis, Georgios and Mouloudakis, Kostas},
  journal = {arXiv:2609.08261},
  year = {2026},
  eprint = {2609.08261},
  archivePrefix = {arXiv},
  primaryClass = {physics.atom-ph},
  url = {https://arxiv.org/abs/2609.08261}
}


\end{document}